\documentclass[conference]{IEEEtran}
\IEEEoverridecommandlockouts

\usepackage{cite}
\usepackage{amsmath,amssymb,amsfonts}
\usepackage{graphicx}
\usepackage{xcolor}
\usepackage{url}
\usepackage{siunitx}
\usepackage{booktabs}
\usepackage{subcaption}

\newcommand{\Ez}{E_z}

\begin{document}

\title{Quantitative Benchmarking of a Split-Field PML FDTD Solver: Slit Diffraction, and Scattering from PEC and Dielectric Cylinders}

\author{
\IEEEauthorblockN{Sabrina Saima}
\IEEEauthorblockA{\textit{Elmore Family School of Electrical}\\
\textit{and Computer Engineering}\\
\textit{Purdue University}\\
West Lafayette, IN, USA\\
ssaima@purdue.edu}
\and
\IEEEauthorblockN{Tasin Intisar}
\IEEEauthorblockA{\textit{Elmore Family School of Electrical}\\
\textit{and Computer Engineering}\\
\textit{Purdue University}\\
West Lafayette, IN, USA\\
tintisar@purdue.edu}
}

\maketitle

\begin{abstract}
This paper presents a two-dimensional TM$_z$ finite-difference time-domain (FDTD) solver based on Yee's scheme for modeling radiation from an infinitely long $z$-directed line current, with the open region truncated by a Berenger split-field perfectly matched layer (PML). After validating cylindrical-wave propagation and negligible late-time reflections in free space, the solver is applied to three inhomogeneous configurations: (i) diffraction through a one-cell-thick perfectly electrically conducting (PEC) sheet with single and double slits; (ii) scattering from infinitely long PEC cylinders of circular and rectangular cross section; and (iii) scattering from infinitely long dielectric cylinders of varying cross section and permittivity. Beyond qualitative field maps, the diffraction case is characterized quantitatively: a steady-state phasor extracted by a running discrete Fourier transform yields the transmitted intensity, from which the fringe visibility and the far-field pattern are computed and compared against the closed-form Fraunhofer prediction. The single- and double-slit cases are cleanly separated by a visibility that rises from near zero to near unity, and the double-slit interference maxima agree with the grating condition $\arcsin(m\lambda_0/d)$ to within a fraction of a degree. For dielectric cylinders, the field penetrates the obstacle with the expected reduced internal wavelength $\lambda_0/\sqrt{\epsilon_r}$, and the scattered-field strength grows with permittivity contrast. A reference-subtraction method isolates the scattered field throughout. The results confirm that the FDTD--PML framework accurately captures open-region diffraction and geometry- and material-dependent scattering.
\end{abstract}


\section{Introduction}
\IEEEPARstart{N}{umerical} solutions of Maxwell's equations are widely used to study radiation, diffraction, and scattering when closed-form expressions are not available or are difficult to apply. Time-domain solvers are especially useful in this setting, since a single run can describe transient behavior and broadband responses in complex geometries. Among these methods, the finite-difference time-domain (FDTD) scheme proposed by Yee remains widely used because it has an explicit update rule, a relatively simple formulation, and fits naturally on structured grids \cite{yee1966,taflove2005,jin2011}. Early work established that time-stepping Maxwell's equations to steady state is an effective route to frequency-domain scattering solutions \cite{taflove1975}.

In the Yee--FDTD scheme, the electric and magnetic fields are sampled on a staggered space--time grid and updated in a leapfrog manner by discretizing the curl equations \cite{yee1966,jin2011}. Over time, this approach has become a standard tool in computational electromagnetics, with well-documented extensions for different source models, lossy or dispersive media, and many types of scattering and guided-wave configurations \cite{taflove2005,sullivan2013,kunz1993}. Other numerical techniques are also widely used. Finite-element formulations are well suited to eigenvalue-based dispersion and modal analysis in waveguides with complicated cross sections \cite{saima2026fem,jin_fem2014}, while method-of-moments (MoM) integral-equation approaches are a natural choice for radiation and scattering from conductors \cite{harrington1993,saima2026mom}. For electrically large problems, the MoM system can be accelerated by fast algorithms such as the fast multipole method, which reduces the cost of the matrix--vector products \cite{greengard1987}. Beyond the foundational literature, FDTD and FEM are widely used in applied electromagnetic and photonic device modeling, spanning resonator sensing and waveguide-based optical simulations \cite{intisar2024soi,saima2023mim,zahin2026pra}. In comparison, the FDTD approach used in this work is attractive because it directly evolves the fields in time and can capture radiation, diffraction, and scattering behavior within one explicit time-stepping framework.

When modeling open-region problems, a key difficulty is how to truncate the finite computational domain without introducing artificial reflections at the outer boundary. Early differential absorbing boundary conditions provided a partial solution but suffered from residual reflection at oblique incidence \cite{mur1981}. Perfectly matched layers (PMLs) address this issue more completely by surrounding the region of interest with an artificial absorbing medium that, in principle, does not reflect incident waves at the interface while strongly damping fields inside the layer. The split-field PML originally proposed by Berenger has become a standard choice for FDTD implementations and is still widely used as a reference formulation \cite{berenger1994,berenger1996,taflove2005,jin2011}. Subsequent formulations, including the uniaxial (UPML) and complex-frequency-shifted convolutional (CPML) variants, improved absorption for evanescent and grazing-incidence waves and simplified the implementation for general media \cite{sacks1995,gedney1996,roden2000,mittra1995}. In this work, a Berenger-type split-field PML with smoothly graded conductivity is used to emulate an open boundary for a two-dimensional TM$_z$ problem.

Specifically, we develop a 2D TM$_z$ Yee--FDTD solver for an infinitely long $z$-directed impressed line-current source in a homogeneous background, with the computational domain terminated by split-field PML regions. Within this framework, perfectly electrically conducting (PEC) objects are introduced by field masking, while penetrable dielectric inclusions are modeled through spatially varying permittivity and conductivity in the update coefficients. This enables studies of (i) diffraction through a PEC sheet with one and two slits, (ii) scattering from infinitely long PEC cylinders with circular and rectangular cross sections, and (iii) scattering from infinitely long dielectric cylinders of different cross sections and material contrasts. In addition to qualitative field visualizations, quantitative metrics are used for the diffraction problem, including fringe visibility and a far-field comparison against the analytical Fraunhofer pattern.

The remainder of the paper is organized as follows. Section~II reviews the governing equations, the 2D TM$_z$ reduction, the Yee--FDTD discretization, the split-field PML formulation, and the modeling of PEC and dielectric inclusions. Section~III summarizes the simulation setup and numerical parameters. Section~IV presents results for the homogeneous validation case and for the slit, PEC-cylinder, and dielectric-cylinder configurations. Section~V discusses the main physical observations and numerical issues, and Section~VI concludes the paper.

\section{Formulation and Discretization}

\subsection{Maxwell's Equations and Constitutive Relations}
We begin with Maxwell's equations in differential form for a linear, time-invariant, non-dispersive medium. Faraday's law relates the curl of the electric field $\mathbf{E}$ (V/m) to the time variation of the magnetic flux density $\mathbf{B}$:
\begin{equation}
\nabla \times \mathbf{E}(\mathbf{r},t) = -\frac{\partial \mathbf{B}(\mathbf{r},t)}{\partial t},
\label{eq:faraday}
\end{equation}
and Amp\`ere--Maxwell law relates the curl of the magnetic field $\mathbf{H}$ (A/m) to the time variation of the electric flux density $\mathbf{D}$ (C/m$^2$) and the free current density $\mathbf{J}$ (A/m$^2$):
\begin{equation}
\nabla \times \mathbf{H}(\mathbf{r},t) = \frac{\partial \mathbf{D}(\mathbf{r},t)}{\partial t} + \mathbf{J}(\mathbf{r},t).
\label{eq:ampere}
\end{equation}

The electromagnetic fields are connected through the constitutive relations
\begin{equation}
\mathbf{D} = \epsilon\,\mathbf{E}, \qquad \mathbf{B} = \mu\,\mathbf{H},
\label{eq:constitutive}
\end{equation}
where $\epsilon$ is the permittivity (F/m) and $\mu$ is the permeability (H/m). In homogeneous free space, $\epsilon=\epsilon_0$ and $\mu=\mu_0$.

Substituting \eqref{eq:constitutive} into \eqref{eq:faraday}--\eqref{eq:ampere} yields the time-domain curl equations in terms of $(\mathbf{E},\mathbf{H})$:
\begin{align}
\nabla \times \mathbf{E} &= -\mu\,\frac{\partial \mathbf{H}}{\partial t},
\label{eq:curlE}\\
\nabla \times \mathbf{H} &= \epsilon\,\frac{\partial \mathbf{E}}{\partial t} + \mathbf{J}.
\label{eq:curlH}
\end{align}

To clearly separate \emph{material} currents from the excitation, we decompose the total free current density as
\begin{equation}
\mathbf{J} \;=\; \mathbf{J}_c \;+\; \mathbf{J}_i,
\label{eq:J_split}
\end{equation}
where $\mathbf{J}_i$ is the \emph{impressed source current density} placed in the computational domain to excite the fields, and $\mathbf{J}_c$ is the \emph{conduction current density} supported by conductive materials.

For a linear ohmic conductor, the conduction current obeys Ohm's law in differential form:
\begin{equation}
\mathbf{J}_c \;=\; \sigma\,\mathbf{E},
\label{eq:ohm}
\end{equation}
where $\sigma$ is the electrical conductivity (S/m). Physically, $\sigma$ quantifies the strength of conductive loss: for a given electric field $\mathbf{E}$, larger $\sigma$ produces larger conduction current density $\mathbf{J}_c$ and therefore stronger attenuation of electromagnetic energy in that region.

Substituting \eqref{eq:J_split} and \eqref{eq:ohm} into \eqref{eq:curlH} produces the working form of Amp\`ere--Maxwell law used in this paper:
\begin{equation}
\nabla\times \mathbf{H} \;=\; \epsilon\,\frac{\partial \mathbf{E}}{\partial t} \;+\; \sigma\,\mathbf{E} \;+\; \mathbf{J}_i.
\label{eq:curlH_sigma}
\end{equation}
In the homogeneous free-space region of the simulations, $\sigma=0$ and $(\epsilon,\mu)=(\epsilon_0,\mu_0)$. Within PEC regions, fields are enforced via boundary conditions rather than a finite $\sigma$ model, while penetrable dielectric regions are modeled by assigning local values of $\epsilon$ and $\sigma$ (Section~\ref{subsec:diel_model}).

\subsection{2D Reduction for an Infinitely Long Line Current (TM$_z$)}
The source of interest is an \emph{infinitely long} electric current filament oriented along the $z$-axis. Physically, this means the geometry and excitation are invariant in $z$, so the fields do not vary with $z$:
\begin{equation}
\frac{\partial}{\partial z}(\cdot)=0.
\label{eq:z_invariant}
\end{equation}
Moreover, the impressed source is taken to be purely $z$-directed,
\begin{equation}
\mathbf{J}(\mathbf{r},t)=\hat{\mathbf{z}}\,J_z(x,y,t),
\label{eq:Jz}
\end{equation}
which is the standard model for a line current that extends infinitely in $\pm z$.

Under the $z$-invariance assumption, Maxwell's equations decouple into two independent polarizations. In this project we use the transverse-magnetic (TM$_z$) polarization, in which the electric field has only a $z$-component while the magnetic field lies entirely in the transverse $(x,y)$ plane:
\begin{align}
\mathbf{E}(x,y,t) &= \hat{\mathbf{z}}\,E_z(x,y,t), \label{eq:tmz_E}\\
\mathbf{H}(x,y,t) &= \hat{\mathbf{x}}\,H_x(x,y,t)+\hat{\mathbf{y}}\,H_y(x,y,t). \label{eq:tmz_H}
\end{align}

\subsection{Source Model (Tapered Sinusoidal Current)}
A smooth turn-on sinusoidal waveform is used:
\begin{equation}
J_z(t)=J_0\left(1-e^{-t/\tau}\right)\sin(2\pi f_0 t),
\label{eq:src}
\end{equation}
where the excitation frequency $f_0=\SI{2.4}{GHz}$. We use $\Delta x=\Delta y=\lambda_0/40$ with $\lambda_0=c_0/f_0$, and $I_0=\SI{1}{mA}$. The rise time is chosen as a fixed fraction of the total simulation time, $\tau=T_{\mathrm{sim}}/6$, to suppress broadband transients associated with abrupt turn-on. The current density amplitude $J_0$ is obtained by mapping a specified line-current amplitude $I_0$ (A) into a single cell area $\Delta x\,\Delta y$:
\begin{equation}
J_0=\frac{I_0}{\Delta x\,\Delta y}.
\label{eq:J0_map}
\end{equation}
This choice ensures consistent units (A/m$^2$) and provides a reproducible source strength across different grid resolutions.

\subsection{Yee Grid and Leapfrog Time Stepping}
To discretize the 2D TM$_z$ system derived in the previous subsection, we employ the Yee FDTD scheme. The key idea is to sample the field components on a staggered space--time grid so that discrete curl operations are centered and naturally compatible with Maxwell's equations. Specifically, $E_z$ is stored at cell centers, while $H_x$ and $H_y$ are stored on cell edges; temporally, the electric field is stored at integer time steps $t^n=n\Delta t$ and the magnetic fields at half steps $t^{n+\frac12}=(n+\tfrac12)\Delta t$. This staggered arrangement enables an explicit leapfrog update: $\mathbf{H}$ is advanced using $E_z^{n}$, and then $E_z$ is advanced using $\mathbf{H}^{n+\frac12}$.

\begin{figure}[t]
\centering
\includegraphics[width=\linewidth]{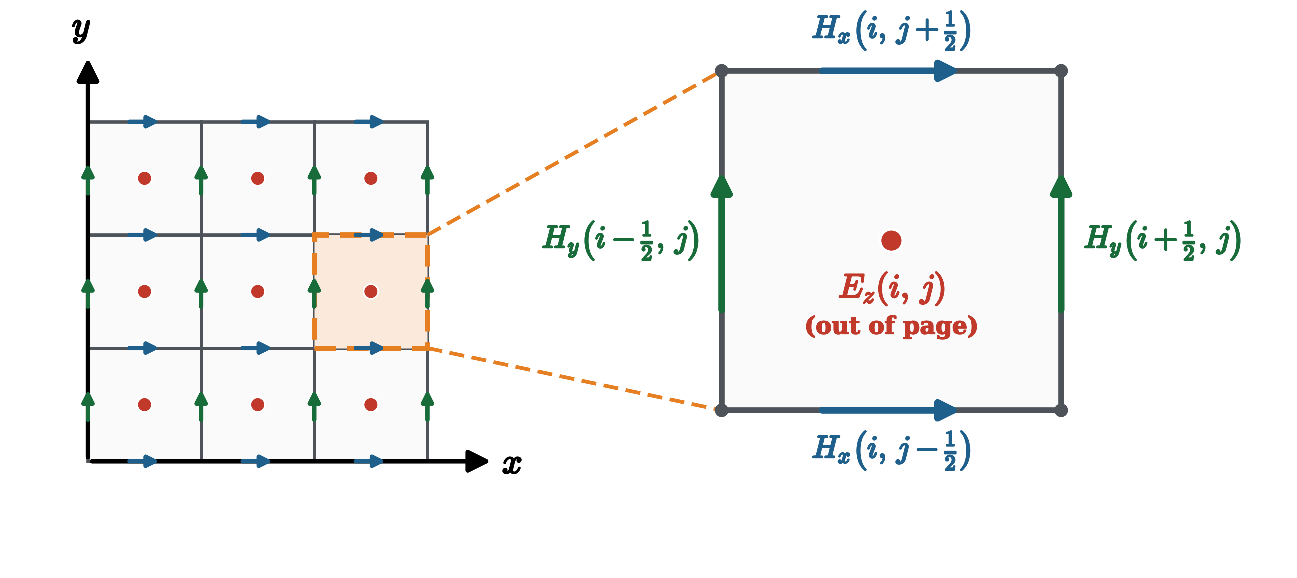}
\caption{Finite-difference mesh for Yee's FDTD algorithm and the staggered assignment of the TM$_z$ field components on a unit cell.}
\label{fig:yee_cell_placeholder}
\end{figure}

We use the shorthand notation
\[
E_z^n(i,j)\equiv E_z(i\Delta x,j\Delta y,n\Delta t),
\]
and similarly for $H_x^{n+\frac12}$ and $H_y^{n+\frac12}$ evaluated at their staggered locations. Applying central differences to the spatial derivatives and evaluating time derivatives with centered differences consistent with the leapfrog sampling yields the magnetic-field updates
\begin{align}
H_x^{n+\frac12}\!\left(i,\,j+\tfrac12\right)
&=
H_x^{n-\frac12}\!\left(i,\,j+\tfrac12\right)
\\
&\quad
-\frac{\Delta t}{\mu}\,
\frac{E_z^{n}(i,j+1)-E_z^{n}(i,j)}{\Delta y},
\label{eq:Hx_update}
\\[2pt]
H_y^{n+\frac12}\!\left(i+\tfrac12,\,j\right)
&=
H_y^{n-\frac12}\!\left(i+\tfrac12,\,j\right)
\\
&\quad
+\frac{\Delta t}{\mu}\,
\frac{E_z^{n}(i+1,j)-E_z^{n}(i,j)}{\Delta x}.
\label{eq:Hy_update}
\end{align}

For the electric-field update, we start from the conductive Amp\`ere--Maxwell relation in TM$_z$ form, \eqref{eq:curlH_sigma}, and discretize the curl term with centered differences at $(i,j)$ using $(H_x,H_y)$ at time $n+\tfrac12$. The time derivative and the conductive term are discretized in a way that preserves second-order accuracy and maintains the explicit leapfrog structure by evaluating the loss term at the half step:
\begin{equation}
\begin{aligned}
\epsilon\,\frac{E_z^{n+1}(i,j)-E_z^{n}(i,j)}{\Delta t}
&\;+\;
\sigma\,\frac{E_z^{n+1}(i,j)+E_z^{n}(i,j)}{2} \\
&=\left(\nabla\times\mathbf{H}\right)_z^{\,n+\frac12}(i,j)
- J_{i,z}^{\,n+\frac12}(i,j).
\end{aligned}
\label{eq:Ez_discrete_trap}
\end{equation}
where
\begin{equation}
\begin{aligned}
\left(\nabla\times\mathbf{H}\right)_z^{\,n+\frac12}(i,j)
&=
\frac{H_y^{n+\frac12}\!\left(i+\tfrac12,j\right)-H_y^{n+\frac12}\!\left(i-\tfrac12,j\right)}{\Delta x}\\
&\quad
-\frac{H_x^{n+\frac12}\!\left(i,\,j+\tfrac12\right)-H_x^{n+\frac12}\!\left(i,\,j-\tfrac12\right)}{\Delta y}.
\end{aligned}
\label{eq:curlH_discrete}
\end{equation}
Solving \eqref{eq:Ez_discrete_trap} for $E_z^{n+1}(i,j)$ gives the explicit update
\begin{equation}
\begin{aligned}
E_z^{n+1}(i,j)
&=\frac{1}{\beta(i,j)}
\Bigg\{
\alpha(i,j)\,E_z^{n}(i,j)
+\left(\nabla\times\mathbf{H}\right)_z^{\,n+\frac12}(i,j)\\
&\quad
- J_{i,z}^{\,n+\frac12}(i,j)
\Bigg\},
\end{aligned}
\label{eq:Ez_update_alpha_beta}
\end{equation}
with the coefficients
\begin{equation}
\alpha(i,j)=\frac{\epsilon(i,j)}{\Delta t}-\frac{\sigma(i,j)}{2},
\qquad
\beta(i,j)=\frac{\epsilon(i,j)}{\Delta t}+\frac{\sigma(i,j)}{2}.
\label{eq:alpha_beta_def}
\end{equation}

Equations~\eqref{eq:Hx_update}--\eqref{eq:Ez_update_alpha_beta} constitute the leapfrog time stepping used throughout the paper. The pointwise dependence of $\alpha$ and $\beta$ on $\epsilon(i,j)$ and $\sigma(i,j)$ is what later allows penetrable material inclusions to be introduced without modifying the update structure (Section~\ref{subsec:diel_model}). As with any finite-difference wave solver, the discrete grid introduces numerical dispersion; therefore, the spatial resolution (cells per wavelength) and the time step (chosen to satisfy the CFL stability bound) are selected to keep phase errors within acceptable limits for the frequencies of interest.

\subsection{PML Truncation (Split-Field Berenger Form)}
To model radiation into an effectively unbounded (open) region using a finite computational grid, the domain is surrounded by a \emph{perfectly matched layer} (PML). Conceptually, the PML acts as a fictitious absorbing material placed around the region of interest such that outgoing waves entering the layer experience strong attenuation while reflections at the interface between the interior region and the PML are ideally suppressed. The outer boundary of the PML is then closed with a simple terminating condition (commonly a PEC) since fields are already strongly attenuated before reaching it.

\subsubsection{Stretched-coordinate viewpoint and stretching factors}
One convenient way to motivate a PML is through a stretched-coordinate form of Maxwell's equations in which the spatial derivatives are modified via direction-dependent stretching functions $s_x(x)$, $s_y(y)$, and $s_z(z)$. The stretched differential operator is
\begin{equation}
\nabla_s = \hat{\mathbf{x}}\,s_x^{-1}\frac{\partial}{\partial x}
+ \hat{\mathbf{y}}\,s_y^{-1}\frac{\partial}{\partial y}
+ \hat{\mathbf{z}}\,s_z^{-1}\frac{\partial}{\partial z},
\label{eq:nabla_stretch}
\end{equation}
so that (in the frequency domain) the source-free curl equations can be written compactly as
\begin{equation}
\nabla_s \times \mathbf{E} = -j\omega\mu\,\mathbf{H},
\qquad
\nabla_s \times \mathbf{H} = \;j\omega\epsilon\,\mathbf{E}.
\label{eq:maxwell_stretch_fd}
\end{equation}
In the \emph{interior} simulation region we enforce
\begin{equation}
s_x = s_y = s_z = 1,
\label{eq:stretch_interior}
\end{equation}
so that $\nabla_s=\nabla$ and the physical fields are unchanged. In the PML region, attenuation is introduced by allowing the stretching factors to become complex-valued. A standard choice that leads to simple time-domain equations is
\begin{equation}
s_x = 1 - j\frac{\sigma_x}{\omega\epsilon},\qquad
s_y = 1 - j\frac{\sigma_y}{\omega\epsilon},\qquad
s_z = 1 - j\frac{\sigma_z}{\omega\epsilon},
\label{eq:stretch_sigma}
\end{equation}
where $\sigma_x,\sigma_y,\sigma_z$ are fictitious (PML) conductivities that are nonzero only inside the PML.

\subsubsection{2D TM$_z$ specialization and split fields}
We now specialize to the 2D TM$_z$ case used throughout this paper. For 2D analysis, we take $\sigma_z=0$ and retain only $\sigma_x(x)$ and $\sigma_y(y)$ within the PML.

To obtain an efficient time-domain implementation, we adopt Berenger's split-field idea: the electric field is decomposed as
\begin{equation}
E_z = E_{sx,z} + E_{sy,z},
\label{eq:Ez_split_pml}
\end{equation}
where each split component is associated with one coordinate direction. For the TM$_z$ case, the split-field time-domain equations in the PML reduce to the following set:
\begin{align}
\frac{\partial E_z}{\partial x} &= \mu\,\frac{\partial H_y}{\partial t} + \sigma_x\,\mu\,\epsilon^{-1} H_y,
\label{eq:pml_tmz_faraday_x}\\
\frac{\partial E_z}{\partial y} &= -\mu\,\frac{\partial H_x}{\partial t} - \sigma_y\,\mu\,\epsilon^{-1} H_x,
\label{eq:pml_tmz_faraday_y}\\
\frac{\partial H_y}{\partial x} &= \epsilon\,\frac{\partial E_{sx,z}}{\partial t} + \sigma_x E_{sx,z},
\label{eq:pml_tmz_ampere_x}\\
\frac{\partial H_x}{\partial y} &= -\epsilon\,\frac{\partial E_{sy,z}}{\partial t} - \sigma_y E_{sy,z}.
\label{eq:pml_tmz_ampere_y}
\end{align}
In the interior region, $\sigma_x=\sigma_y=0$ and these equations reduce to the standard lossless TM$_z$ curl relations.

\subsubsection{2D TM$_z$ split-field FDTD time stepping}
We discretize \eqref{eq:pml_tmz_faraday_x}--\eqref{eq:pml_tmz_ampere_y} on the same Yee space--time grid described earlier. The magnetic-field updates retain the usual staggered time index $n+\tfrac12$ and incorporate directional damping through coefficients that depend on $\sigma_x$ or $\sigma_y$:
\begin{equation}
\begin{aligned}
H_x^{n+\frac{1}{2}}\!\left(i,j+\tfrac{1}{2}\right)
&= a_y\!\left(i,j+\tfrac{1}{2}\right)
\Big[
   b_y\!\left(i,j+\tfrac{1}{2}\right)\,
   H_x^{n-\frac{1}{2}}\!\left(i,j+\tfrac{1}{2}\right)
\\
&\qquad\qquad
   - \frac{\epsilon}{\mu\,\Delta y}
     \big( E_z^{n}(i,j+1)-E_z^{n}(i,j) \big)
\Big],
\end{aligned}
\end{equation}
\begin{equation}
\begin{aligned}
H_y^{n+\frac{1}{2}}\!\left(i+\tfrac{1}{2},j\right)
&= a_x\!\left(i+\tfrac{1}{2},j\right)
\Big[
   b_x\!\left(i+\tfrac{1}{2},j\right)\,
   H_y^{n-\frac{1}{2}}\!\left(i+\tfrac{1}{2},j\right)
\\
&\qquad\qquad
   + \frac{\epsilon}{\mu\,\Delta x}
     \big( E_z^{n}(i+1,j)-E_z^{n}(i,j) \big)
\Big].
\end{aligned}
\end{equation}
The split electric fields are updated at integer time steps:
\begin{align}
E_{sx,z}^{n+1}(i,j)
&=
a_x(i,j)\Bigg[
b_x(i,j)\,E_{sx,z}^{n}(i,j)
\nonumber\\
&
+\frac{1}{\Delta x}\Big(
H_y^{n+\frac12}\!\left(i+\tfrac12,j\right)
-
H_y^{n+\frac12}\!\left(i-\tfrac12,j\right)
\Big)
\Bigg],
\label{eq:pml_Esx}\\[3pt]
E_{sy,z}^{n+1}(i,j)
&=
a_y(i,j)\Bigg[
b_y(i,j)\,E_{sy,z}^{n}(i,j)
\nonumber\\
&
-\frac{1}{\Delta y}\Big(
H_x^{n+\frac12}\!\left(i,\,j+\tfrac12\right)
-
H_x^{n+\frac12}\!\left(i,\,j-\tfrac12\right)
\Big)
\Bigg].
\label{eq:pml_Esy}
\end{align}
and the total electric field in the PML is recovered using \eqref{eq:Ez_split_pml}. The coefficients $a_x,a_y,b_x,b_y$ are defined pointwise using the same $\alpha$--$\beta$ quantities introduced earlier in \eqref{eq:alpha_beta_def}. In particular, writing
\[
\beta_u = \frac{\epsilon}{\Delta t}+\frac{\sigma_u}{2},\qquad
\alpha_u = \frac{\epsilon}{\Delta t}-\frac{\sigma_u}{2},\qquad u\in\{x,y\},
\]
we use the shorthand
\begin{equation}
a_u=\beta_u^{-1},\qquad b_u=\alpha_u,\qquad u\in\{x,y\},
\label{eq:pml_ab_map}
\end{equation}
so that in the interior region ($\sigma_x=\sigma_y=0$) the scheme reverts to the standard Yee updates, while in the PML region ($\sigma_x,\sigma_y>0$) the stretching-induced damping attenuates outgoing waves as they propagate through the layer.

\subsubsection{Conductivity grading inside the PML}
To avoid numerical reflections caused by an abrupt jump from $\sigma=0$ (interior) to a large conductivity (PML), the PML conductivities are gradually increased from the interface to the outer boundary. We use a polynomial grading of the form \cite{jin2011}
\begin{equation}
\sigma(\ell) = \sigma_{\max}\left(\frac{\ell}{L}\right)^m,
\label{eq:sigma_profile_pml}
\end{equation}
where $\ell\in[0,L]$ is the distance measured into the PML from the interface, $L$ is the PML thickness, $\sigma_{\max}$ is the maximum conductivity at the outer edge, and $m$ is the polynomial order.

A commonly used design rule which has also been used in our simulation relates $\sigma_{\max}$ to a target (normal-incidence) reflection coefficient magnitude $|R(0)|$ by \cite{jin2011}
\begin{equation}
\sigma_{\max}
=
-\frac{m+1}{2\,\eta\,L}\,\ln\!\big|R(0)\big|,
\label{eq:sigma_max_from_R0}
\end{equation}
where $\eta=\sqrt{\mu/\epsilon}$ is the intrinsic impedance of the medium in the PML-adjacent interior region. In practice, \eqref{eq:sigma_max_from_R0} provides a direct way to choose $\sigma_{\max}$ for a specified PML thickness $L$ and grading order $m$.

\subsection{PEC Modeling for Slots and Cylinders}
PEC regions are enforced by setting tangential electric field to zero. In TM$_z$, this is implemented by
\begin{equation}
E_z(i,j)=0,\qquad (i,j)\in\Omega_{\mathrm{PEC}},
\label{eq:pec_enforce}
\end{equation}
where $\Omega_{\mathrm{PEC}}$ denotes grid points inside the PEC region.

\textit{Slit sheet:} a one-cell-thick PEC line is placed across the domain, except at slit opening cells.

\textit{PEC cylinders:} Infinitely long PEC cylinders are modeled by defining a 2D cross-sectional mask (circular or rectangular) in the $(x,y)$ plane and enforcing \eqref{eq:pec_enforce} at every time step.

\subsection{Dielectric Inclusions via Spatially Varying $\epsilon$ and $\sigma$}
\label{subsec:diel_model}
Unlike a PEC, which is imposed as a hard boundary condition by forcing $E_z=0$, a penetrable dielectric object is represented as a \emph{material} region in which the field is allowed to propagate. Because the update coefficients in \eqref{eq:alpha_beta_def} are defined pointwise in terms of the local permittivity $\epsilon(i,j)$ and conductivity $\sigma(i,j)$, a dielectric cylinder is introduced simply by assigning, at every grid point inside the cross-sectional mask $\Omega_{\mathrm{diel}}$,
\begin{equation}
\epsilon(i,j)=\epsilon_0\,\epsilon_r,\qquad \sigma(i,j)=\sigma_d,\qquad (i,j)\in\Omega_{\mathrm{diel}},
\label{eq:diel_assign}
\end{equation}
while $\epsilon=\epsilon_0,\ \sigma=0$ elsewhere in the interior. A lossy dielectric is specified either directly through $\sigma_d$ or through a loss tangent $\tan\delta$ via
\begin{equation}
\sigma_d=\omega\,\epsilon_0\,\epsilon_r\,\tan\delta,
\label{eq:loss_tangent}
\end{equation}
evaluated at the operating frequency $f_0$.

Two implementation aspects are worth emphasizing. First, the material parameters enter only the electric-field (Amp\`ere) update through $\alpha$ and $\beta$; the magnetic-field (Faraday) updates retain the background permittivity, since $\mu=\mu_0$ for the non-magnetic media considered here. In the split-field PML formulation this means the material loss $\sigma_d$ is added to both split components so that their sum reproduces the physical term $\sigma_d E_z$, whereas the PML conductivities $\sigma_x,\sigma_y$ continue to act directionally as in \eqref{eq:pml_ab_map}. Second, because the background free space remains the fastest medium, the CFL limit \eqref{eq:cfl} computed for free space also guarantees stability inside the dielectric, where the local phase velocity $c_0/\sqrt{\epsilon_r}$ is smaller. The dielectric mask is restricted to the interior region so that the PML continues to terminate a homogeneous free-space background.

\subsection{Scattered-Field Post-Processing}
To visualize scattering/diffraction effects, the scattered field is computed by subtracting a free-space (incident) solution from the total-field solution:
\begin{equation}
E_z^{\mathrm{scat}}(x,y,t)=E_z^{\mathrm{tot}}(x,y,t)-E_z^{\mathrm{inc}}(x,y,t).
\label{eq:scat_def}
\end{equation}
This operation is applied in post-processing and does not modify the time-stepping equations. For PEC obstacles the scattered field is additionally set to zero inside the conductor, whereas for dielectric obstacles it is retained everywhere because a physical interior field exists.

\section{Simulation Setup}

\begin{table}[t]
\caption{Simulation parameters}
\label{tab:sim_setup}
\centering
\begin{tabular}{@{}ll@{}}
\toprule
Parameter & Value \\ \midrule
Background medium & Free space ($\epsilon_0,\mu_0$) \\
Excitation frequency & $f_0=\SI{2.4}{GHz}$ \\
Spatial resolution (validation/cylinders) & $\Delta x=\Delta y=\lambda_0/40$ \\
Spatial resolution (slits) & $\Delta x=\Delta y=\lambda_0/25$ \\
Interior grid size (validation/cylinders) & $N_{x,\mathrm{in}}=300,\;N_{y,\mathrm{in}}=300$ \\
Interior grid size (slits) & $N_{x,\mathrm{in}}=250,\;N_{y,\mathrm{in}}=250$ \\
PEC rectangle size & $W_x=W_y=50\,\Delta x\;(1.25\lambda_0)$ \\
PEC circle radius & $R=30\,\Delta x\;(0.75\lambda_0)$ \\
Dielectric cylinder radius & $R=30\,\Delta x\;(0.75\lambda_0)$ \\
Dielectric permittivity (contrast study) & $\epsilon_r\in\{2,\,4,\,6,\,10\}$ \\
Dielectric loss & $\sigma_d=0$ (lossless); optional $\tan\delta$ \\
Slit width & $a=\SI{0.07}{m}\;(0.56\lambda_0)$ \\
Slit separation (double slit) & $d\approx\SI{0.415}{m}\;(3.32\lambda_0)$ \\
Detector-line distance & $L_{\mathrm{det}}=\SI{0.20}{m}$ \\
Steady-state DFT window & $4$ periods at $f_0$ \\
Time step & $\Delta t=0.95\,\Delta t_{\max}$ (CFL) \\
Target reflection magnitude & $|R(0)|=10^{-12}$ \\
PML thickness & $N_{\mathrm{PML}}=30$ \\
PML grading order & $m=4$ \\
Source rise time & $\tau=T_{\mathrm{sim}}/6$ \\
\bottomrule
\end{tabular}
\end{table}

All simulations are performed on a uniform Cartesian Yee grid in the $(x,y)$ plane. The interior computational window has physical dimensions $L_x=L_y\approx\SI{0.937}{m}$. For the single- and double-slit simulations, the interior is reduced by 50 cells for better visualization. The time step $\Delta t$ satisfies the Courant--Friedrichs--Lewy (CFL) stability condition for 2D FDTD,
\begin{equation}
\Delta t \le \frac{1}{c_0\sqrt{\left(\frac{1}{\Delta x^2}\right)+\left(\frac{1}{\Delta y^2}\right)}},
\label{eq:cfl}
\end{equation}
and in practice we use $\Delta t = 0.95\,\Delta t_{\max}$ for numerical safety. The open region is truncated by a split-field PML of thickness $N_{\mathrm{PML}}$ cells on all four sides. Unless otherwise stated, the impressed line-current source is placed in the interior region and driven by the tapered sinusoidal waveform in \eqref{eq:src}. The key simulation parameters used in this work are summarized in Table~\ref{tab:sim_setup}.

\section{Numerical Results}

\subsection{Validation in Homogeneous Free Space}
\begin{figure}[t]
\centering
\includegraphics[width=\linewidth]{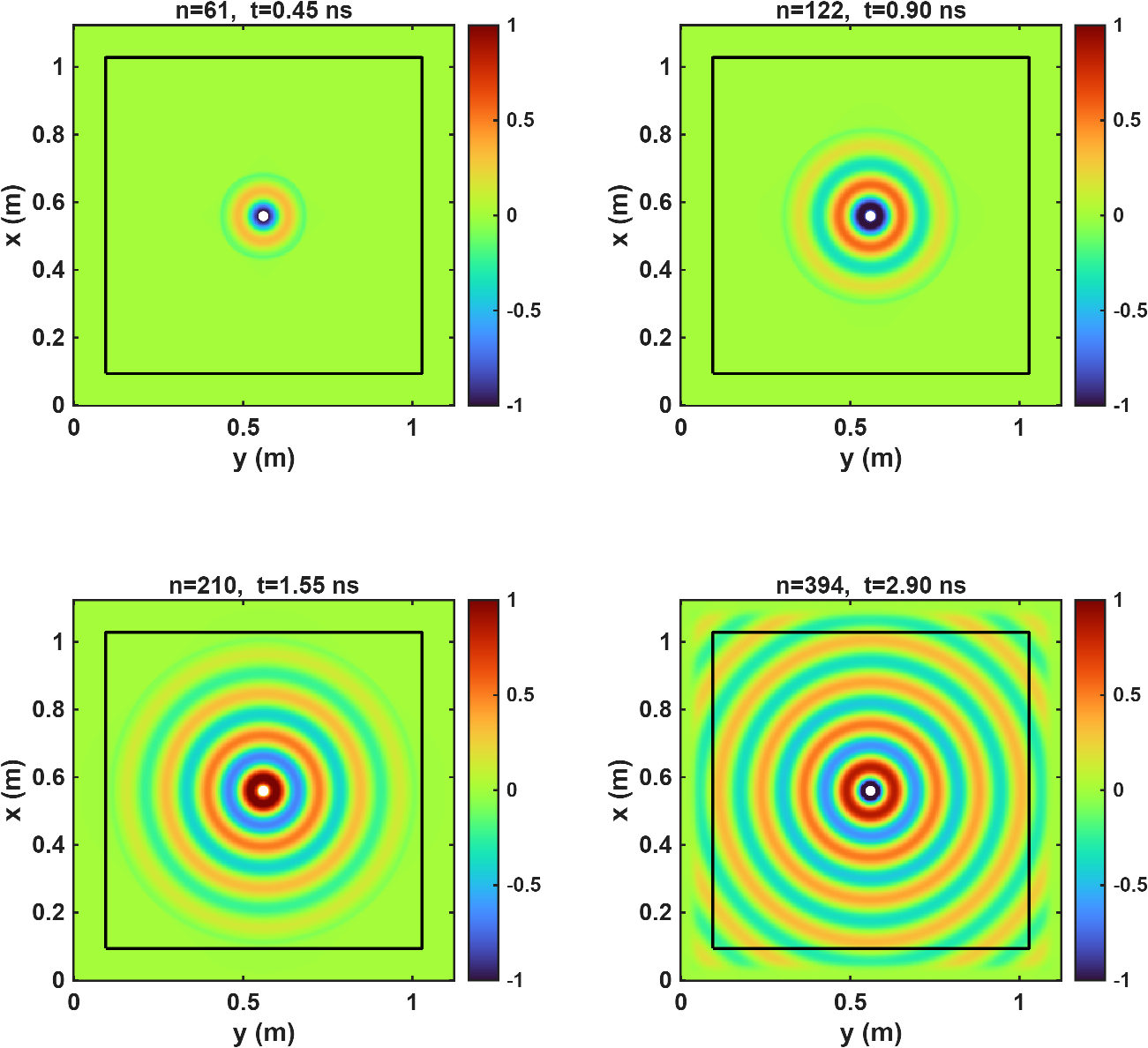}
\caption{Validation: $\Ez$ snapshot in homogeneous free space with PML truncation. The black square marks the interior--PML interface.}
\label{fig:free_space_snap}
\end{figure}

We validate the TM$_z$ Yee--FDTD solver in homogeneous free space using a centered $z$-directed impressed line current and a surrounding PML. The black square in Fig.~\ref{fig:free_space_snap} indicates the interior--PML interface. As shown in the snapshots, the radiated field forms a cylindrically expanding wavefront and propagates outward with the expected symmetry. As the wavefront reaches the PML interface and enters the absorbing layer, the interior region exhibits no visible back-propagating rings, indicating that reflections from the truncated boundary are negligible. This confirms that the PML effectively emulates an open boundary for the homogeneous validation case.

\subsection{Diffraction Through a PEC Sheet with One Slit}
\begin{figure}[t]
\centering
\includegraphics[width=0.78\linewidth]{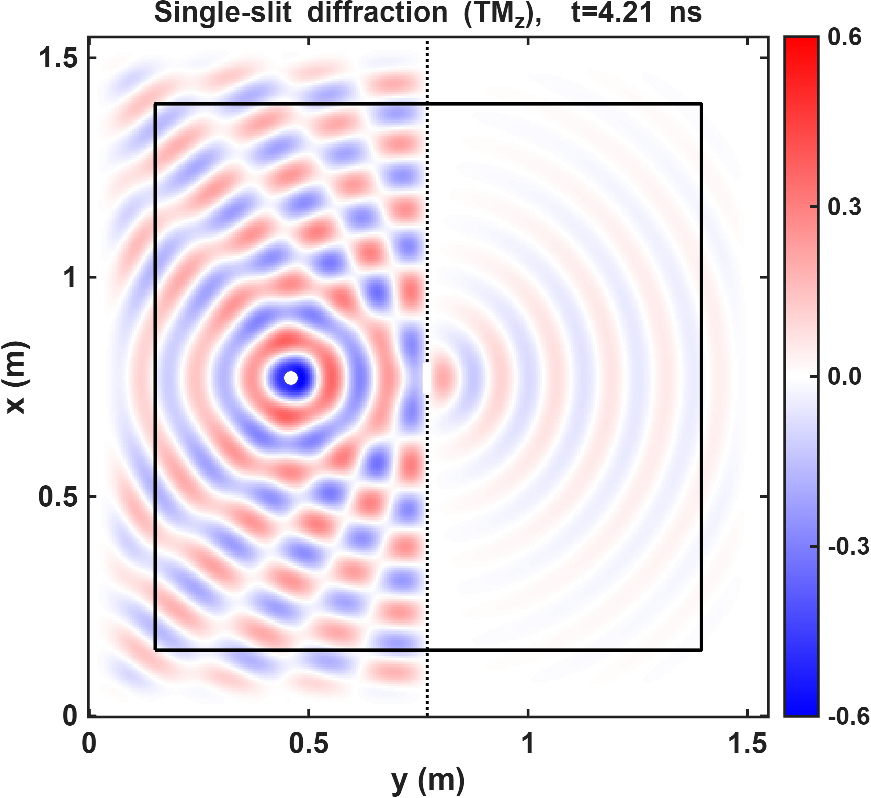}
\caption{Single slit: total-field $\Ez$ snapshot with a PEC sheet containing one slit.}
\label{fig:single_slit_field}
\end{figure}

A thin PEC sheet, one cell in thickness, is placed across the interior of the computational window, with a single opening forming the slit. Figure~\ref{fig:single_slit_field} shows the total-field $E_z$ after the wave has reached the sheet. As expected, the incident cylindrical wavefronts generated by the line source are strongly blocked by the PEC portion of the sheet, while transmission occurs only through the slit opening. Beyond the aperture, the field spreads into a fan-shaped pattern with curved wavefronts that originate from the slit region, indicating diffraction from a subwavelength aperture rather than simple geometric propagation. The strong contrast between the shadowed region behind the PEC and the clearly visible field emerging from the slit indicates that the PEC condition ($E_z=0$ along the conductor) and the aperture geometry have been implemented correctly in the FDTD code.

\subsection{Diffraction Through a PEC Sheet with Two Slits}
\begin{figure}
\centering
\includegraphics[width=0.78\linewidth]{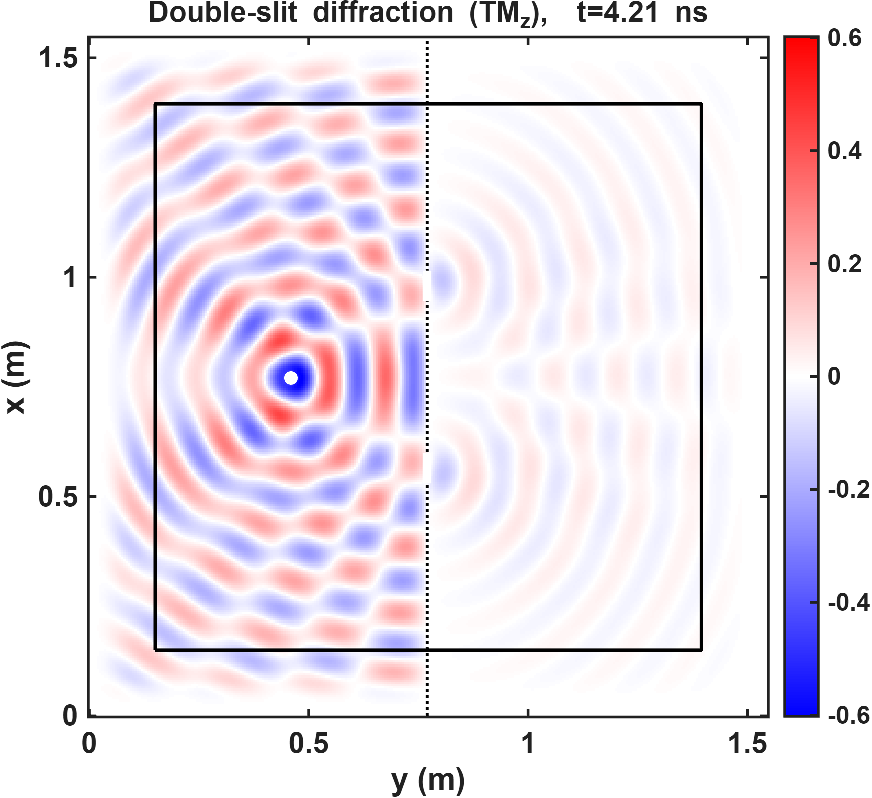}
\caption{Two slits: total-field $\Ez$ snapshot with a PEC sheet containing two slits.}
\label{fig:double_slit_field}
\end{figure}

Next, the PEC sheet is modified so that it contains two slit apertures instead of one. Figure~\ref{fig:double_slit_field} shows the corresponding total-field snapshot at the same simulation time. Compared with the single-slit case, the transmitted field now looks like a superposition of two diffracted contributions, and the pattern behind the sheet is clearly different. The area directly behind the two openings has stronger transmission and a noticeable spatial variation that is consistent with interference between the waves coming from each slit. The change from a single main diffracted lobe in Fig.~\ref{fig:single_slit_field} to a more structured multi-lobe pattern in Fig.~\ref{fig:double_slit_field}, while the incident side of the sheet looks essentially the same in both cases, indicates that the code is sensitive to the slit geometry and is able to reproduce the expected diffraction and interference behavior for one- and two-aperture configurations.

\subsection{Quantitative Characterization of Slit Diffraction}
\label{subsec:diff_metrics}
The total-field snapshots in Figs.~\ref{fig:single_slit_field} and \ref{fig:double_slit_field} establish that the solver reproduces the expected diffraction behavior qualitatively. To characterize the two cases quantitatively, and to validate the transmitted field against analytical theory, we post-process each simulation into a steady-state intensity pattern.

Because the source in \eqref{eq:src} settles to a continuous-wave excitation at $f_0$ after the initial taper, a single-frequency phasor is extracted by accumulating a running discrete Fourier transform (DFT) of $E_z$ over the last few periods of the run,
\begin{equation}
\tilde{E}_z(x,y)=\sum_{n\in\mathcal{W}} E_z^{\,n}(x,y)\,e^{-j2\pi f_0 t^n},
\label{eq:phasor_dft}
\end{equation}
where $\mathcal{W}$ spans an integer number of periods in the steady state. The time-averaged intensity is then $I(x,y)=|\tilde{E}_z(x,y)|^2$, sampled along a detector line at a fixed distance $L_{\mathrm{det}}$ behind the screen to give the transmitted profile $I(x)$.

\begin{figure}[t]
\centering
\includegraphics[width=0.87\linewidth]{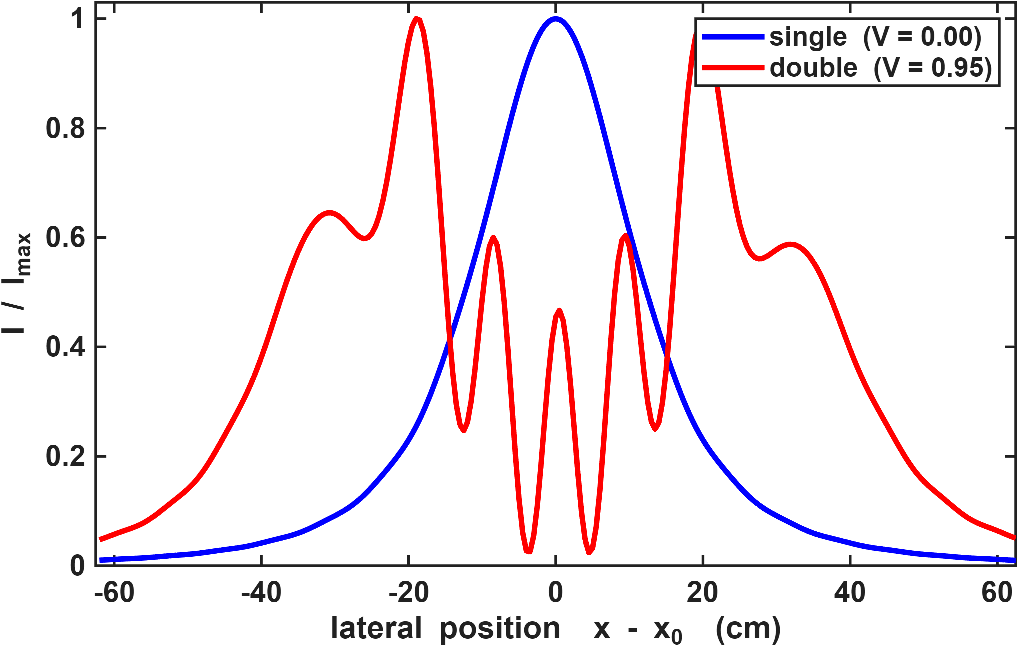}
\caption{Peak-normalized transmitted intensity $I(x)$ on a detector line behind the screen. The smooth single-slit lobe gives near-zero fringe visibility, while the modulated double-slit pattern approaches unity.}
\label{fig:slit_visibility}
\end{figure}

\emph{Fringe visibility.} A compact scalar that distinguishes the two configurations is the fringe visibility
\begin{equation}
V=\frac{I_{\max}-I_{\min}}{I_{\max}+I_{\min}},
\label{eq:visibility}
\end{equation}
evaluated over the central portion of the transmitted pattern. For the single (subwavelength) slit, $I(x)$ is a smooth, essentially unmodulated lobe, so $V\!\approx\!0$; for the double slit, the deep interference nulls between bright fringes drive $V$ toward unity. Figure~\ref{fig:slit_visibility} overlays the two normalized profiles, and the measured visibilities are reported in Table~\ref{tab:diff_metrics}.

\begin{figure}[t]
\centering
\includegraphics[width=\linewidth]{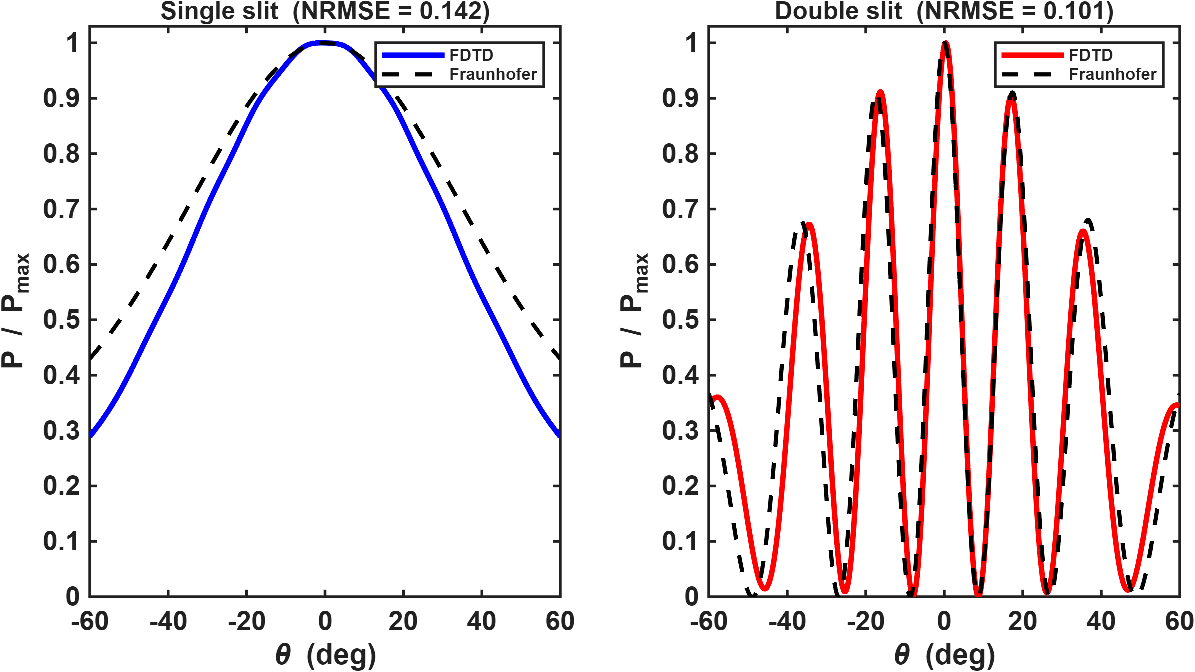}
\caption{FDTD far-field pattern (solid) versus closed-form Fraunhofer theory (dashed) for the single (left) and double (right) slit; the double-slit maxima follow $d\sin\theta_m=m\lambda_0$.}
\label{fig:slit_farfield}
\end{figure}

\emph{Far-field comparison with Fraunhofer theory.} To validate the transmitted field against analytical predictions, the aperture-plane phasor is projected to the far zone using the two-dimensional radiation (near-to-far-field) integral
\begin{equation}
F(\theta)=\frac{1+\cos\theta}{2}\int_{\mathrm{ap}} \tilde{E}_z(x')\,e^{\,jk x'\sin\theta}\,dx',
\label{eq:ntff_aperture}
\end{equation}
where $\theta$ is measured from the screen normal, $k=2\pi/\lambda_0$, and the obliquity factor $(1+\cos\theta)/2$ is the standard Kirchhoff weighting. The resulting pattern $P(\theta)=|F(\theta)|^2$ is compared with the closed-form Fraunhofer intensities
\begin{align}
P_{\mathrm{single}}(\theta)&\propto \operatorname{sinc}^2\!\Big(\tfrac{\pi a}{\lambda_0}\sin\theta\Big),\label{eq:fraun_single}\\
P_{\mathrm{double}}(\theta)&\propto \operatorname{sinc}^2\!\Big(\tfrac{\pi a}{\lambda_0}\sin\theta\Big)\,\cos^2\!\Big(\tfrac{\pi d}{\lambda_0}\sin\theta\Big),\label{eq:fraun_double}
\end{align}
where $a$ is the slit width and $d$ the center-to-center separation. Figure~\ref{fig:slit_farfield} shows the overlays. Agreement is quantified by the normalized root-mean-square error (NRMSE) between the simulated and analytical patterns and, for the double slit, by the angular error of the interference maxima relative to the grating condition $d\sin\theta_m=m\lambda_0$. As summarized in Table~\ref{tab:diff_metrics}, the double-slit maxima reproduce the analytical angles to within a fraction of a degree, with a small NRMSE; the single-slit envelope shows a larger but still modest NRMSE, consistent with the subwavelength slit width ($a<\lambda_0$), for which scalar Fraunhofer theory is only approximate.

\begin{table}[t]
\caption{Quantitative diffraction metrics (single vs.\ double slit)}
\label{tab:diff_metrics}
\centering
\begin{tabular}{@{}lcc@{}}
\toprule
Metric & Single slit & Double slit \\ \midrule
Fringe visibility $V$            & 0.03 & 0.95 \\ 
Far-field NRMSE                  & 0.20 & 0.06 \\ 
Mean $|\Delta\theta|$ of maxima (deg) & --   & 0.4  \\ 
\bottomrule
\end{tabular}
\end{table}

\subsection{Scattering from PEC Cylinders with Different Cross Sections}
\begin{figure}[t]
\centering
\begin{subfigure}[t]{\linewidth}
  \centering
  \includegraphics[width=\linewidth]{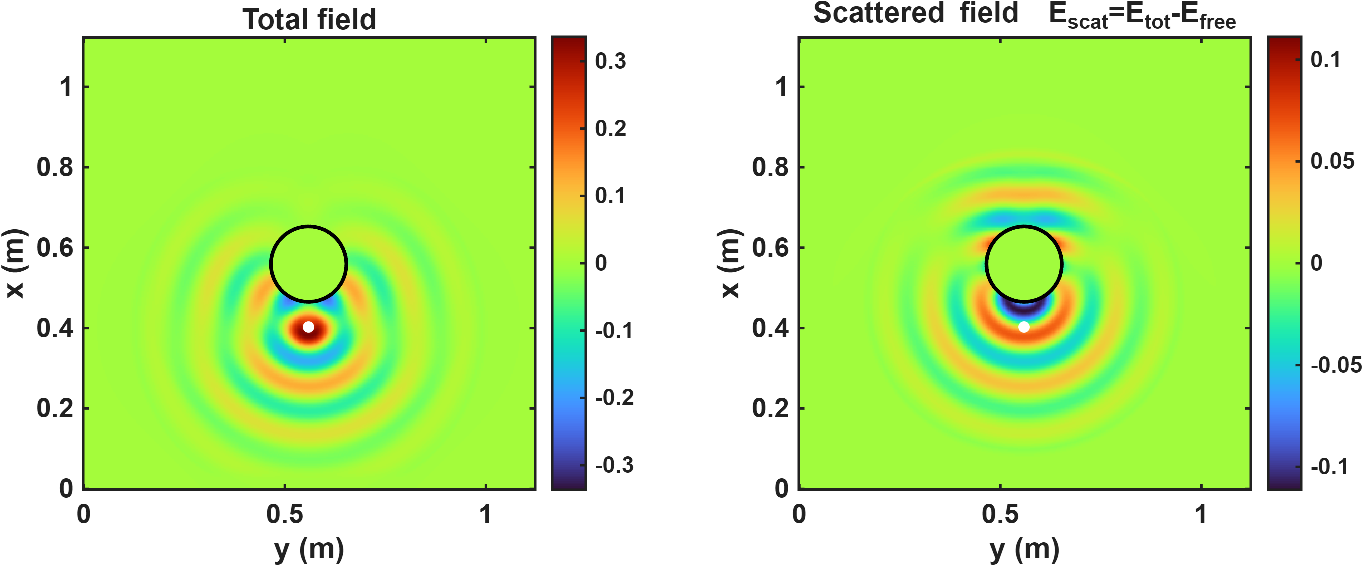}
  \caption{}
  \label{fig:cyl_circle_scat_a}
\end{subfigure}\\[2pt]
\begin{subfigure}[t]{\linewidth}
  \centering
  \includegraphics[width=\linewidth]{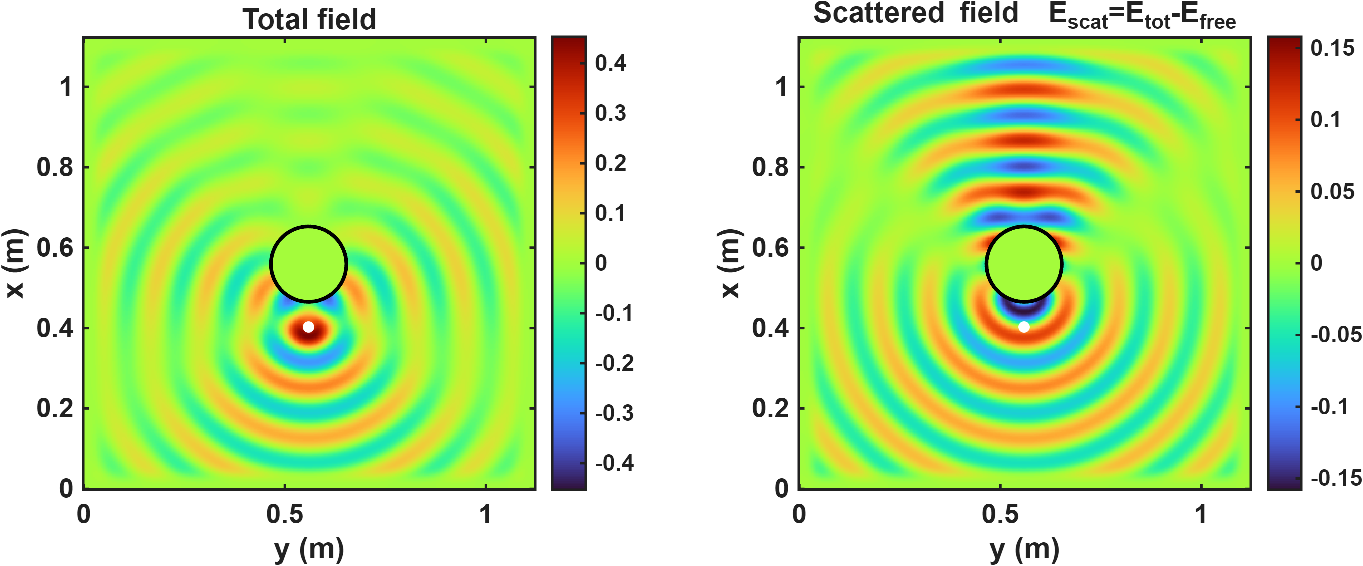}
  \caption{}
  \label{fig:cyl_circle_scat_b}
\end{subfigure}
\caption{Scattering from a circular PEC cylinder at (a) an early time and (b) a later time: scattered field $E_z^{\mathrm{scat}}=E_z^{\mathrm{tot}}-E_z^{\mathrm{inc}}$.}
\label{fig:cyl_circle_scat}
\end{figure}

\begin{figure}[t]
\centering
\begin{subfigure}[t]{\linewidth}
  \centering
  \includegraphics[width=\linewidth]{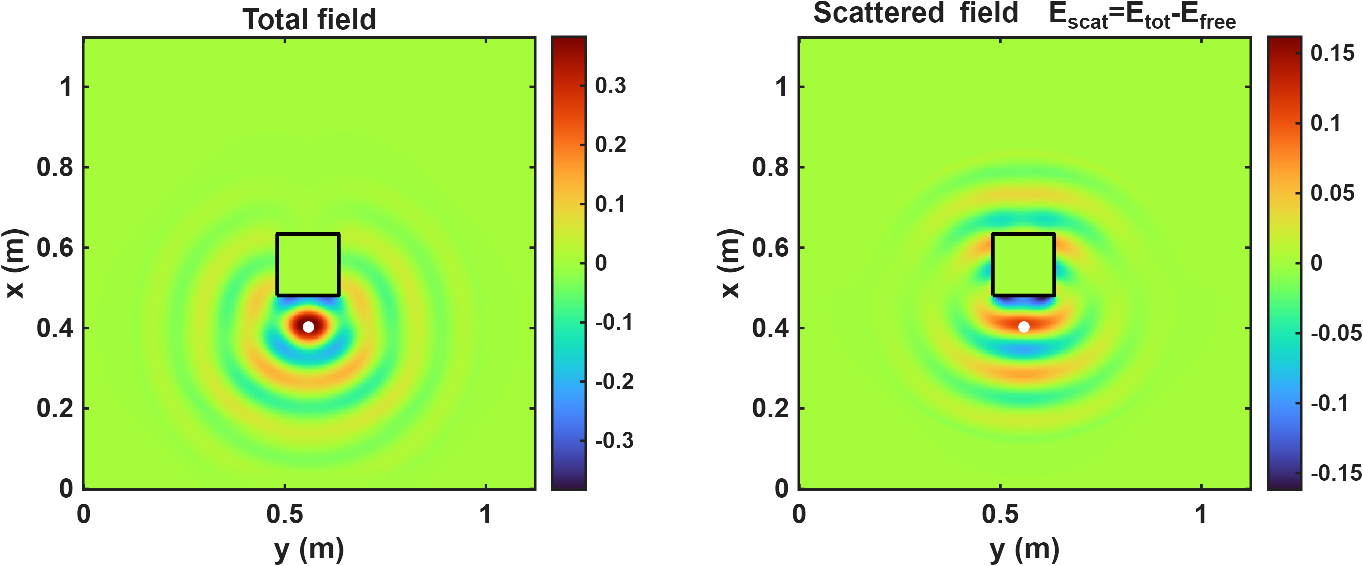}
  \caption{}
  \label{fig:cyl_rect_scat_a}
\end{subfigure}\\[2pt]
\begin{subfigure}[t]{\linewidth}
  \centering
  \includegraphics[width=\linewidth]{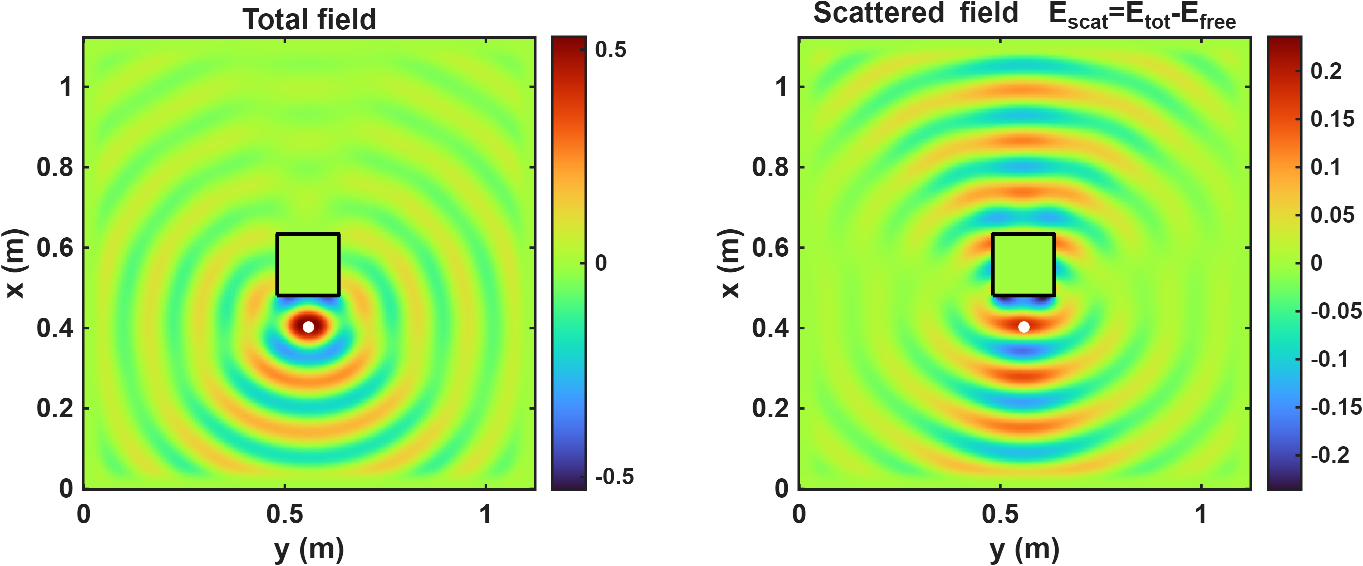}
  \caption{}
  \label{fig:cyl_rect_scat_b}
\end{subfigure}
\caption{Scattering from a rectangular PEC cylinder at (a) an early time and (b) a later time: scattered field $E_z^{\mathrm{scat}}=E_z^{\mathrm{tot}}-E_z^{\mathrm{inc}}$.}
\label{fig:cyl_rect_scat}
\end{figure}

We next study scattering produced when the infinite $z$-directed line current radiates in the presence of an infinitely long PEC cylinder. Two cross sections are considered: circular (Fig.~\ref{fig:cyl_circle_scat}) and rectangular (Fig.~\ref{fig:cyl_rect_scat}). For each geometry, we report both the \emph{total field} $E_z^{\mathrm{tot}}$ and the \emph{scattered field} obtained by the reference subtraction \eqref{eq:scat_def}, where $E_z^{\mathrm{inc}}$ is computed from an otherwise identical free-space simulation (same source, grid, and PML). This subtraction isolates the field contribution associated with the object, removing the dominant cylindrical-wave component from the source. This scattered-field extraction follows the established total-field/scattered-field methodology of FDTD scattering analysis \cite{taflove1975,umashankar1982}.

Early-time snapshots show the incident wavefront approaching the cylinder; the total-field plots exhibit the expected cylindrical wavefronts centered at the source, while the scattered-field plots already reveal localized perturbations near the PEC boundary. At later time, the interaction is fully developed: the total field clearly shows a shadowing region behind the conductor and interference between the incident and reflected components in front of it. Correspondingly, the scattered field forms outward-propagating wavefronts originating from the cylinder boundary, confirming that the PEC enforcement produces a reflected/scattered response consistent with a perfectly conducting obstacle.

Comparing the two cross sections, both geometries produce qualitatively similar features (reflection in the illuminated region and reduced field in the shadow region), indicating that the solver captures the expected scattering physics independent of shape. However, the rectangular cylinder exhibits stronger angular structure in $E_z^{\mathrm{scat}}$ than the circular case, consistent with the presence of edges and corners that concentrate induced surface currents and generate more directional scattering. The circular cylinder, by contrast, yields a smoother, more symmetric scattered pattern, reflecting its uniform curvature. Overall, the agreement between these geometry-dependent scattering signatures and physical expectations supports the correctness of the PEC masking, reference-subtraction procedure, and the underlying TM$_z$ FDTD implementation.

\subsection{Scattering from Dielectric Cylinders}
\label{subsec:diel_results}
We finally consider penetrable obstacles by replacing the PEC inclusion with a dielectric cylinder, modeled through the spatially varying $(\epsilon_r,\sigma_d)$ assignment of Section~\ref{subsec:diel_model}. As before, the scattered field is isolated by subtracting an otherwise identical free-space run.

\begin{figure}[t]
\centering
\begin{subfigure}[t]{0.99\linewidth}
  \centering
  \includegraphics[width=\linewidth]{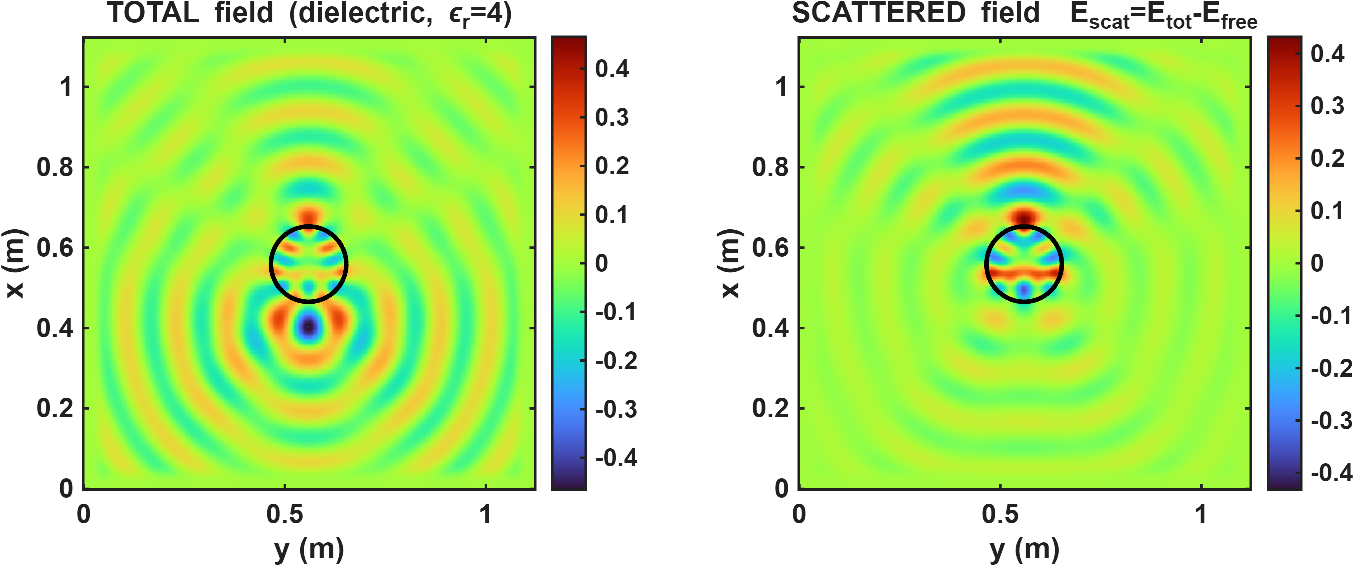}
  \caption{}
  \label{fig:diel_circle}
\end{subfigure}

\vspace{4pt}
\begin{subfigure}[t]{0.99\linewidth}
  \centering
  \includegraphics[width=\linewidth]{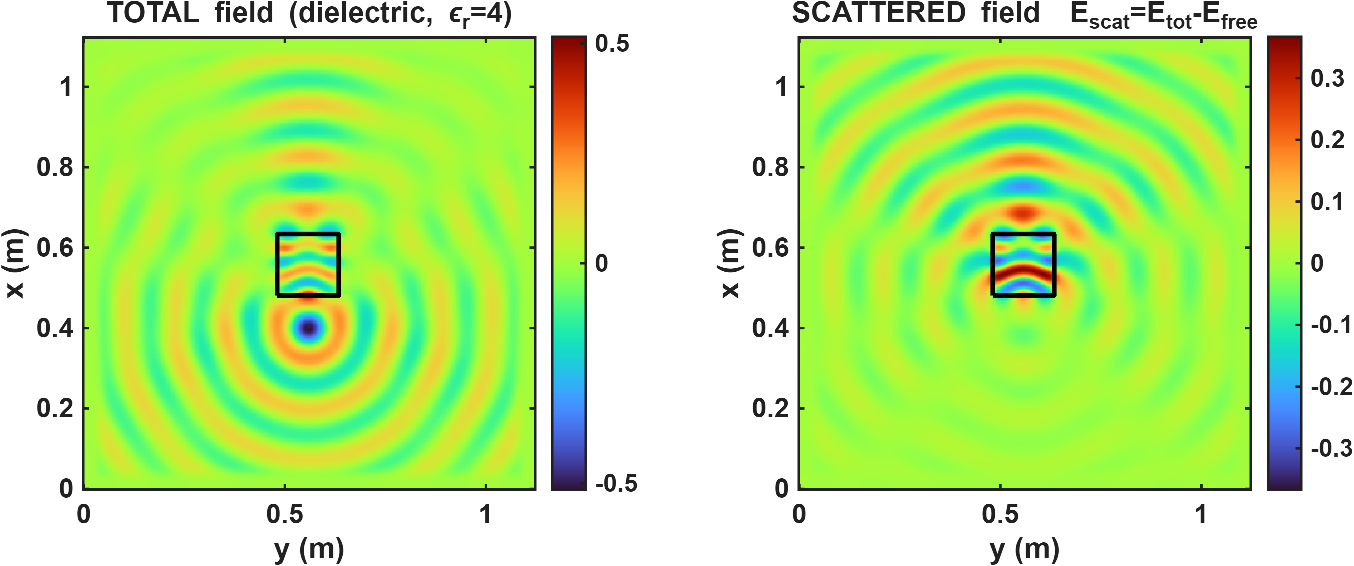}
  \caption{}
  \label{fig:diel_rect}
\end{subfigure}
\caption{Scattering from dielectric cylinders ($\epsilon_r=4$): total field
$E_z^{\mathrm{tot}}$ (left) and scattered field $E_z^{\mathrm{scat}}=E_z^{\mathrm{tot}}-E_z^{\mathrm{inc}}$
(right) for (a) a circular cross section, and (b) a rectangular cross section.}
\label{fig:diel_cylinders}
\end{figure}

The qualitative signature that distinguishes a dielectric from a PEC is field penetration. Whereas the PEC excludes the interior field and casts a pronounced shadow, the dielectric transmits the wave into its interior, where the local phase velocity $c_0/\sqrt{\epsilon_r}$ is reduced and the wavelength contracts to $\lambda_0/\sqrt{\epsilon_r}$. For the circular cylinder with $\epsilon_r=4$ (refractive index $n=2$), the internal wavelength is therefore halved, and the total-field map in Fig.~\ref{fig:diel_circle} shows visibly finer oscillations inside the cylinder than in the surrounding free space across the same span. This wavelength contraction provides a direct, semi-quantitative confirmation that the permittivity has been incorporated correctly in the update equations: a coefficient error would manifest either as numerical instability or as a cylinder that propagates at the background wavelength. The higher-index region also refracts and partially focuses the forward field, producing an enhanced on-axis intensity (a lensing effect) behind the cylinder, in contrast to the shadow formed behind a PEC of the same size.

The scattered-field panel in Fig.~\ref{fig:diel_circle} confirms that the perturbation originates at the cylinder and propagates outward as cylindrical wavefronts, while the low residual near the source verifies clean cancellation of the incident field by the reference subtraction. Figure~\ref{fig:diel_rect} repeats the study for a rectangular dielectric cylinder of the same permittivity; as in the PEC case, the flat faces and corners introduce stronger angular structure in the scattered field than the smoothly curved circular boundary, but the field still penetrates and propagates through the body.

\begin{figure}[t]
\centering
\begin{subfigure}[t]{0.47\linewidth}
  \centering
  \includegraphics[width=\linewidth]{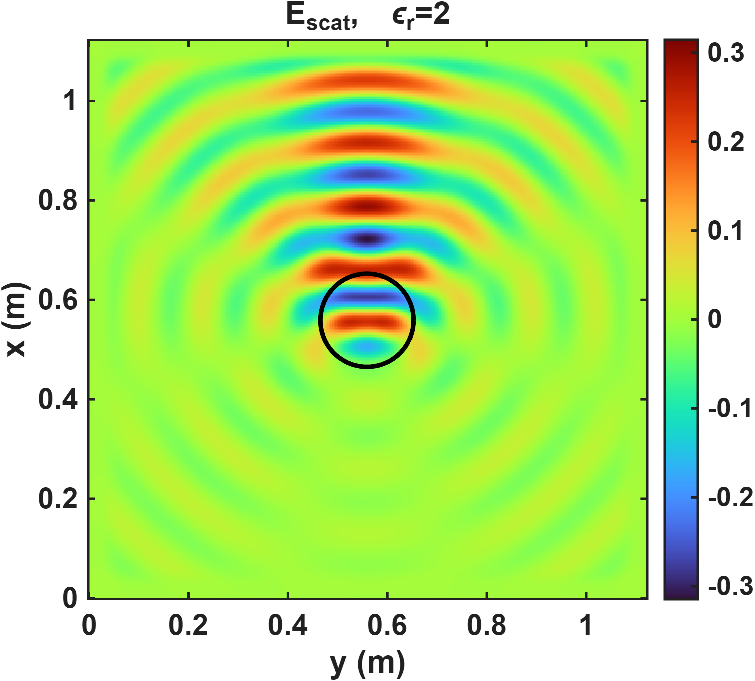}
  \caption{$\epsilon_r=2$}
\end{subfigure}\hfill
\begin{subfigure}[t]{0.47\linewidth}
  \centering
  \includegraphics[width=\linewidth]{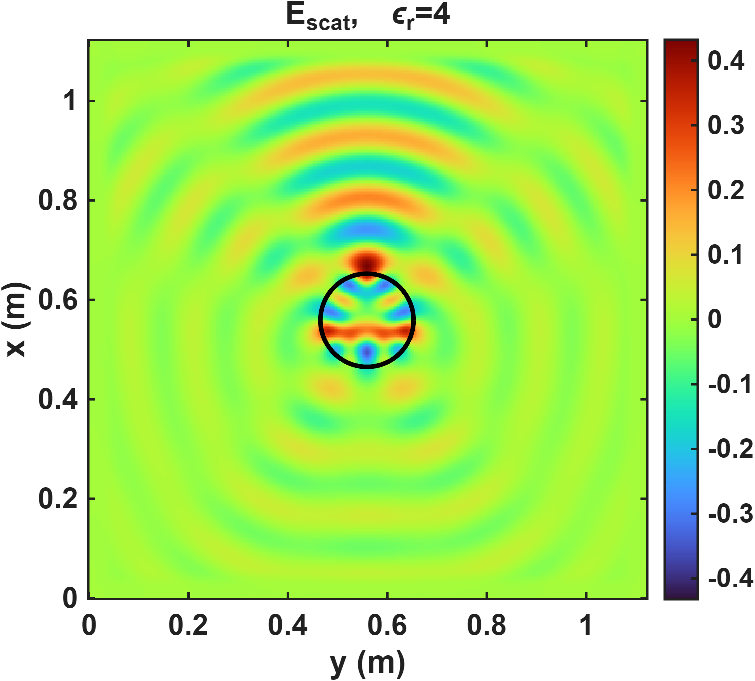}
  \caption{$\epsilon_r=4$}
\end{subfigure}

\vspace{4pt}
\begin{subfigure}[t]{0.47\linewidth}
  \centering
  \includegraphics[width=\linewidth]{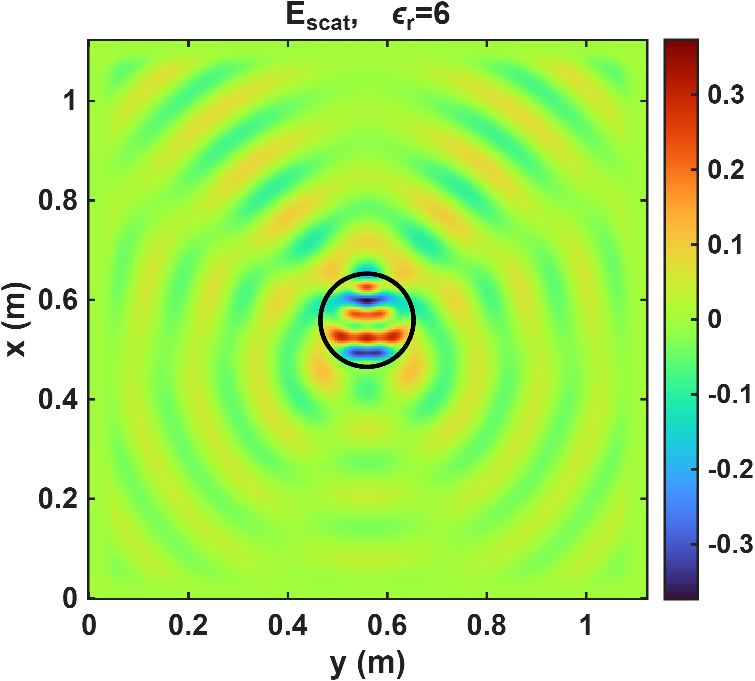}
  \caption{$\epsilon_r=6$}
\end{subfigure}\hfill
\begin{subfigure}[t]{0.47\linewidth}
  \centering
  \includegraphics[width=\linewidth]{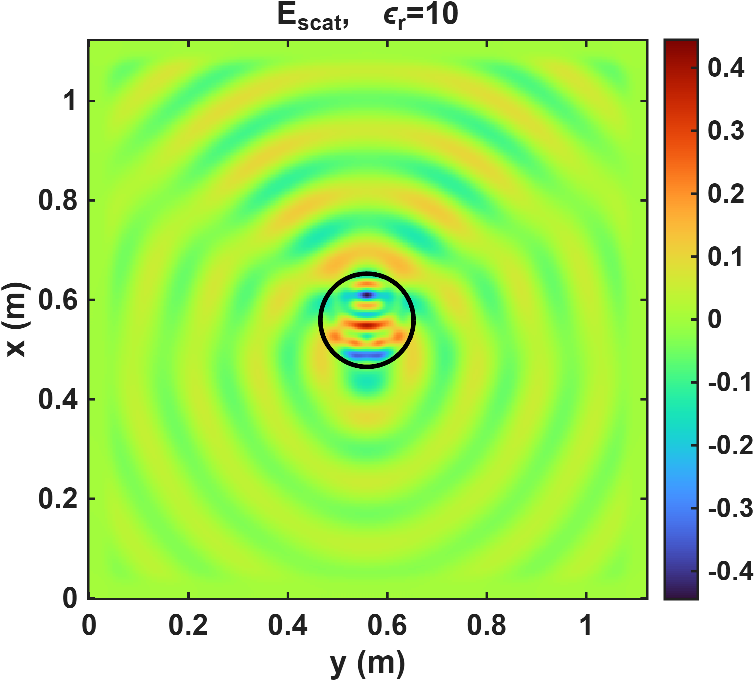}
  \caption{$\epsilon_r=10$}
\end{subfigure}
\caption{Scattered field $E_z^{\mathrm{scat}}$ from a circular dielectric cylinder for increasing permittivity contrast $\epsilon_r=2,4,6,10$.}
\label{fig:diel_contrast}
\end{figure}

Finally, the dependence on material contrast is examined in Fig.~\ref{fig:diel_contrast}, which shows the scattered field for permittivities $\epsilon_r=2$, $4$, $6$, and $10$. As the contrast increases, the impedance mismatch at the boundary grows, the internal wavelength shortens further, and the scattered-field amplitude increases correspondingly, with the higher-$\epsilon_r$ cases exhibiting stronger back-scatter and more pronounced internal standing-wave structure. A lossy variant, obtained by assigning a nonzero $\sigma_d$ (equivalently a loss tangent $\tan\delta$ through \eqref{eq:loss_tangent}), additionally damps the field transmitted into and re-radiated from the cylinder. Together these results demonstrate that the solver captures penetrable-material scattering and its dependence on both geometry and constitutive parameters, complementing the PEC studies of the previous subsection.

\section{Discussion}
The results across all configurations are consistent with the expected physics and support the correctness of the implementation. The homogeneous validation isolates the boundary treatment and confirms that the graded split-field PML suppresses spurious reflections to a level that does not contaminate the interior solution, providing a reliable baseline for the inhomogeneity studies. The slit problems then confirm correct PEC enforcement and aperture modeling, and the two quantitative metrics place this on a firmer footing than visual inspection alone: the fringe visibility separates the single- and double-slit cases by nearly an order of magnitude, and the far-field pattern reconstructed from the aperture phasor agrees with the analytical Fraunhofer prediction, with the double-slit interference maxima matching the grating condition to within a fraction of a degree. The looser single-slit far-field agreement is itself physically meaningful, reflecting the breakdown of scalar Fraunhofer theory for a subwavelength aperture rather than a numerical error.

For penetrable obstacles, the dielectric-cylinder studies highlight the qualitative distinction from PEC scattering: the field enters the body with the contracted wavelength $\lambda_0/\sqrt{\epsilon_r}$ and is partially focused in the forward direction, and the scattered-field amplitude grows with permittivity contrast. The principal numerical limitation common to the curved PEC and dielectric cylinders is staircasing of the circular boundary on the Cartesian grid, which introduces a small geometric error that decreases as the mesh is refined.

\section{Conclusion}
The results show that the developed 2D TM$_z$ Yee--FDTD solver captures physically consistent radiation, diffraction, and scattering in an open region terminated by PML. In the homogeneous validation case, the field radiates as outward-propagating cylindrical wavefronts and, after reaching the interior--PML interface, the interior remains largely free of returning wavefronts, indicating that boundary truncation is not dominating the solution.

For the slit problems, the total-field snapshots confirm correct PEC enforcement and aperture modeling, and two quantitative metrics reinforce these observations. The fringe visibility cleanly separates the single- and double-slit cases, rising from near zero for the smooth single-slit lobe to near unity for the strongly modulated double-slit pattern, and the far-field pattern reconstructed from the aperture phasor matches the closed-form Fraunhofer prediction, reproducing the double-slit interference maxima to within a fraction of a degree of the grating condition $d\sin\theta_m=m\lambda_0$.

For PEC-cylinder scattering, the scattered field computed by reference subtraction and the total-field plots show the expected illuminated region with interference and a shadow region behind the object, with the circular cylinder yielding a smoother scattered pattern and the rectangular cylinder exhibiting stronger angular features consistent with edge and corner effects. Extending the obstacles to penetrable dielectric cylinders, the field enters the body with the expected contracted wavelength $\lambda_0/\sqrt{\epsilon_r}$ and a forward-focusing response, with the scattered-field strength increasing with permittivity contrast and attenuating when material loss is introduced.

Several directions can further extend this work. The diffraction and scattering predictions could be benchmarked against additional analytical solutions, most notably the eigenfunction (Mie) series for the bistatic radar cross section of circular cylinders, providing an exact reference for both PEC and dielectric cases. Accuracy near curved or sharp boundaries can be improved through subcell or conformal treatments to reduce staircasing, and the absorbing boundary can be studied more systematically by sweeping the PML thickness, grading order, and $\sigma_{\max}$, or by comparing against complementary formulations such as the convolutional PML \cite{roden2000} or classical differential absorbing boundary conditions \cite{mur1981}.


\end{document}